# NOAA Image Data Acquisition to Determine Soil Moisture in Arequipa-Perú

A. Argume, R. Coaguila, P.R. Yanyachi and J. Chilo

*Abstract*—In recent years, irrigations have been built on dry areas in Majes-Arequipa. Over time, the irrigations water forms moist areas in lower areas, which can have positive or negative consequences. Therefore, it is important to know in advance where the water from the new irrigation will appear. The limited availability of real-time satellite image data is still a hindrance to some applications. Data from NOAA's environmental satellites are available fee and license free. In order to receive data, users must obtain necessary equipment. In this work we present a satellite data acquisition system with an RTL SDR receiver, two 137-138 Mhz designed antennas, Orbitron, SDRSharp, WXTolmag and MatLab software. We have designed two antennas, a Turnstile Crossed dipole antenna with Balun and a quadrifilar helicoidal antenna. The antennas parameter measurements show very good correspondence with those obtained by simulation. The RTL SDR RTL2832U receiver, combined with our antennas and software, forms the system for recording, decoding, editing and displaying Automatic Picture Transmission (APT) signals. The results show that the satellite image receptions are sufficiently clear and descriptive for further analysis.

## I. INTRODUCTION

THE world's population is constantly increasing. We expect that in a few years: the food deficit will decrease, the food consumption per capita will increase in poor countries and that the dietary habits of the population will diversify. All these changes will affect food production systems, natural resources and the environment. To meet this challenge, it is necessary to increase food production and natural resources. In recent decades, irrigations have been built in the southern part of Peru. The water that is abundant in the mountain area is directed on canals to the arid zones of the coast. One of these irrigations is located in the town Majes-Arequipa. After a few years of operation, the appearance of humid areas in the lower parts has been discovered. It would be very important to predict the location of these areas using satellite images.

Satellites provide information on cloud formations and movements, precipitation, temperatures, ocean currents, sea surface temperatures, air and water pollution, droughts and floods, severe weather conditions, vegetation, insect infestation, atmospheric ozone content, volcanic eruptions and other factors affecting our daily lives. Automatic Picture Transmission (APT) services are designed to transmit live satellite imagery to low cost ground reception stations. A ground station consists of a cheap controllable directional antenna or fixed directional antenna, a VHF receiver, software and a personal computer [1].

Information about soil moisture is important in garden and agricultural environments, various activities such as planting time, irrigation needs and others can be planned better. The authors in [2] have developed an estimation method of live fuel moisture content using NOAAs data: ratio of Normalized Difference Vegetation Index (NDVI), surface temperature (Ts), and the relative greenness (RGRE). In reference [3], soil moisture, evaporation and runoff were estimated using one-layer hydrological model, which takes observed precipitation and temperature data.

NOAA and agencies [4] provide a suite of soil moisture products ranging from modelled estimates over broad areas to actual measurements at specific locations. These products are not available in all parts of the world and in many cases it is necessary to build own ground station for specific applications. There are commercial integrated ground stations that we can buy for some thousand dollars. These systems include satellite tracking, longitude / latitude recording, temperature calibration and scheduled data entry. Developments in electronics and software in recent years have made it possible to design and build cheaper direct-reading ground stations. In this work we have built a satellite image data acquisition system using the RTL SDR RTL2832U that is cheap and easy to install. An Turnstile Crossed dipole antenna was designed, with a Balun to improve the properties of the antenna. Additionally a quadrifilar helicoidal antenna was design. Orbitron, SDRSharp and WXtomag software were used to detect the NOAA satellite, record the satellite signals and decode them. We also use Matlab for subsequent processing of images.

This paper is organized as follows. Section II outlines the methods and materials used and describes the satellite image data acquisition system. Section III presents the results and analysis. Finally, Section IV provides some conclusions.

A. Argume is with Universidad Privada de Tacna-Perú (aargume@gmail.com).

R. Coaguila is with Universidad Católica de Santa María, Arequipa-Perú (rcoaguilag@unsa.edu.pe).

P.R. Yanyachi are with Universidad Nacional San Agustin de Arequipa-Perú (raulpab@unsa.edu.pe).

J. Chilo is with the Electronics Engineering Department, University of Gävle, Sweden (jco@hig.se).

## II. NOAA IMAGE DATA ACQUISITION SYSTEM

The APT signal is continuously transmitted from the satellites. This results in an image strip that continues as long as the transmission is received at the ground station. Radio reception of the APT signal is limited to "line of sight" from the ground station and can therefore only be received when the satellite is above the horizon of a user's ground station. This is determined by both the height of the satellite and its special path over the ground station reception area [1].

A ground station typically contains the following components: antenna, preamplifier, radio receiver, demodulator, software to predict when the satellite will be in view of the ground station and software to record and process the image.

### A. Antennas

Directional antennas and higher gain beam antennas can be properly designed to provide adequate APT reception when used with a good preamplifier and radio receiver. Meteorological satellites such as NOAA use clockwise circular polarization (right) [5], so a Turnstile antenna or a quadrifilar helicoidal antenna is optimal for this purpose, given its radiation pattern and the circular polarization characteristic. In this work we have design and fabricate both antennas.

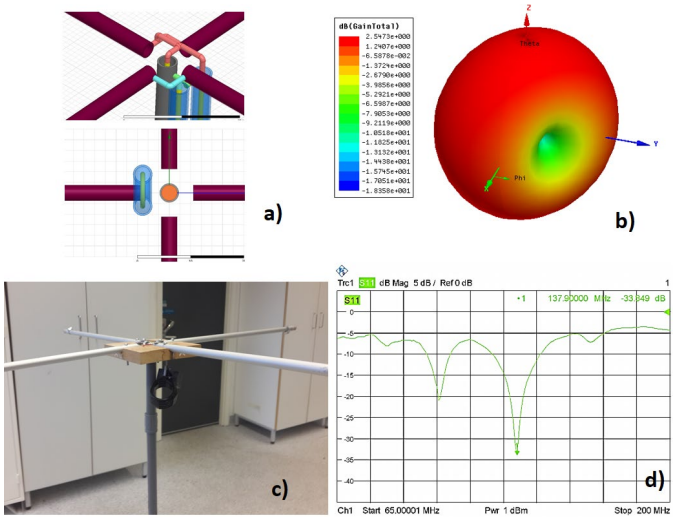

Fig. 1. HFSS simulation a); 3D radiation pattern b); the antenna with Balun c); and S11 measurements using VNA d).

Turnstile antenna is fabricate using two dipole antennas that are placed perpendicular to each other and the current / voltage that interpolates them at a 90 degree phase will produce helical polarization [6]. This antenna is omnidirectional, with little or any signal gain and due to the design and placement of the antenna elements, the satellite signal is often received better in one direction than others [1].

Our Turnstile antenna was designed for VHF bandwidth of 137-138 MHz, consisting of a pair of dipoles (10mm / 2mm) perpendicular to each other. The antenna length was determined to be 1.04 meters. The antenna is improved with the use of a Balun that cancels interferences and reflections [7]. The

Balun is designed with a coaxial cable RG58 with a length of $\lambda / 2 = 0.86$ m. Figure 1 shows the results of antenna simulation in HFSS software with 3D radiation pattern, the antenna constructed and the measurements of S11 using VNA. The antenna S11 parameter measurements show very good correspondence with those obtained by simulation.

A quadrifilar helicoidal antenna consists of four 1/2-turn helices equally spaced around the circumference of a common cylinder. The radiation pattern is omnidirectional in the plane perpendicular to its main axis. Radiation of the signal is nearly circularly polarized over the entire hemisphere irradiated. Authors in [8] and [9] have used these antennas showing a god receive signal quality. Figure 2 presents the results of quadrifilar helicoidal antenna simulation in HFSS software with 3D radiation pattern, the antenna constructed and the simulations of Gain and S11.

The dimensions of the quadrifilar helicoidal antenna are calculated using equation (1) to (5).

The length of helical components of a half loop, L, is determined by [10]:

$$L + 2r = f_1 \frac{\lambda}{2},\qquad(1)$$

$$L^2 = L_{ax}^2 + (2\pi n r)^2,\qquad(2)$$

$$r = \frac{RL_{ax}}{2},\qquad(3)$$

$$L_{ax} = \frac{f_1 \lambda}{2[\sqrt{(n\pi R)^2 + 1}]} + \frac{1}{R},\qquad(4)$$

$$L = f_1 \frac{\lambda}{2} - RL_{ax},\qquad(5)$$

where $f_1$ is the frequency, $R$ is the radius of cylinder, $n$ is number of turns in the helix.

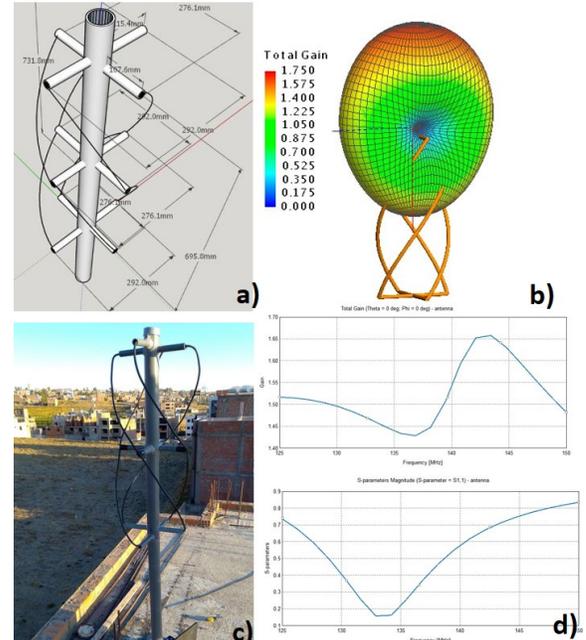

Fig. 2. HFSS simulation a); 3D radiation pattern b); the quadrifilar helicoidal antenna c); and Gain and S11 simulations d).

## B. RTL SDR Receiver

The RTL SDR RTL2832U receiver supports software-defined radio (SDR) mode and communicate over a USB-bus. Main characteristics of the receiver are as follows: 8 bits analog-to digital converter, frequency range from 100 kHz to 2000MHz and maximum sampling frequency of 3.2 MHz. RTL-SDR receivers main advantages are extremely low price, small size and wide frequency range.

The satellite signals from the antenna to the RTL-SDR receiver must exceed acceptable limits, therefore the components of the transmission system must be properly designed to avoid signal loss [11]. We have been careful to choose coaxial cable from antenna to receiver and have made measurements to ensure good signal quality. Figure 3, upper figure, shows the process of receiving satellite signals, where the hardware part and the software part are required. Figure 3, lower figure, shows the IC of the RTL SDR RTL2832U.

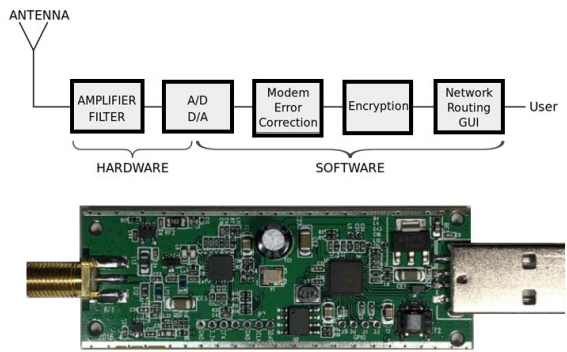

Fig. 3. Receiving satellite signals steps (upper) and IC of the RTL SDR RTL2832U (lower).

## C. Software

A tracker is needed to find out where the satellites are in the polar orbit and predict when they will pass at a sufficiently high angle. In this work we use the software Orbitron.

Orbitron uses a set of numbers for each satellite, Two Line Elements (TLE), to determine where the satellite will be. The software comes with satellite lists and their TLE numbers and it is possible to create custom lists for the satellites we are interested. In Figure 4 we can see the polar orbit for NOAA 15 and NOAA 18 in South America.

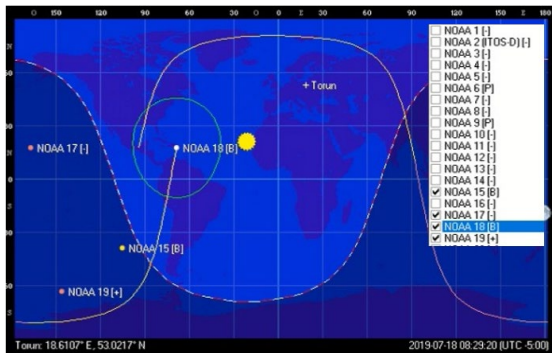

Fig. 4. NOAA 15 and NOAA 18 position in South America.

By setting the satellite altitude to a number between 30 and 45 degrees, the software shows the satellites that are high enough in the sky. After calculations, a list is created with dates and times that show when the satellites will be at their highest altitude.

SDRsharp software is required to control RTL SDR RTL2832U. In SDRsharp, each satellite broadcasts at a slightly different frequency, for example NOAA 15 broadcasts at 137.63 MHz.

For recording settings, broadband FM (WFM) and audio must be selected. It should be noted that the signal bandwidth is about 50 kHz, but it is recommended to set the SDR bandwidth wider than that. When satellite comes over the horizon and we receive radio signals, SDRsharp shows wave tops that leave a very clear line on the waterfall, see Figure 5.

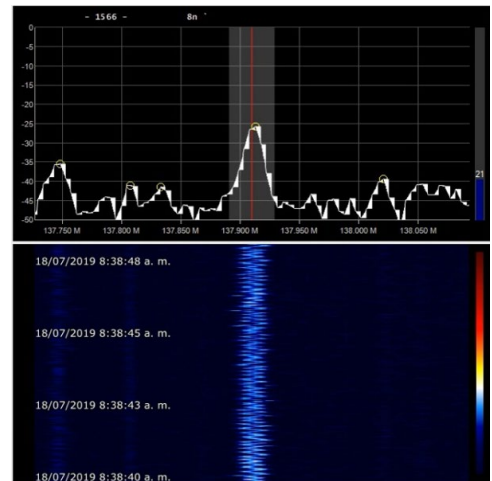

Fig. 5. SDRsharp signal waterfall.

The WXtoIMG software is used to convert the radio signal we recorded into an image. The audio file is loaded into the WXtoIMG software, which automatically decodes the audio file and produces an image. Two different images are displayed, a much darker image which is the visible channel and a brighter image which is the infrared channel. The two channels can combine in different ways to improve the results. By entering the GPS location of the base station, the software will add a map to the image. Figure 6 shows the results from WXtoIMG software.

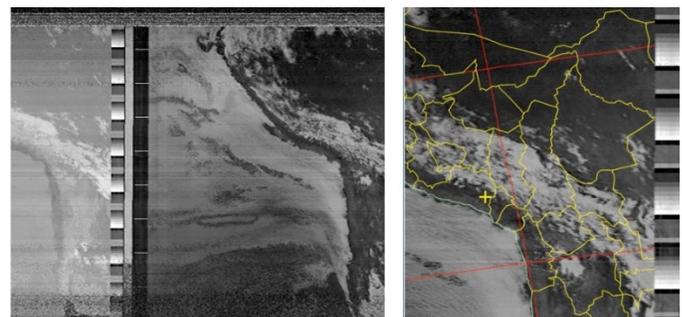

Fig. 6. WXtoIMG two different images (left) and the image with added map (right).

### III. NOAA 15 SATELLITE IMAGES OVER PERÚ

NOAA-15 was launched May 13, 1998, into a 450-nmi (833-km) morning orbit. NOAA 15 was the first in the ATN series to support dedicated microwave instruments for the generation of temperature, moisture, surface, and hydrological products [12].

Figure 7 shows the path of NOAA 15 through the Peruvian sky, indicating the times at certain places.

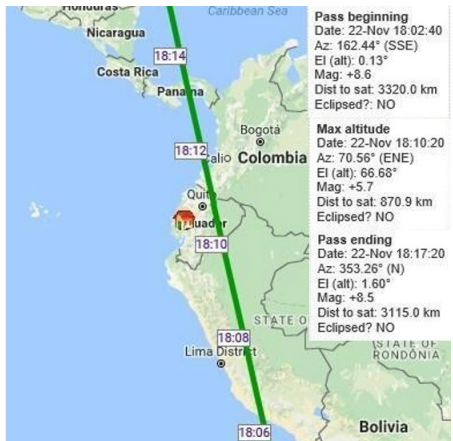

Fig. 7. NOAA 15 path through the Peruvian sky.

Figure 8 presents the results obtained in our project. We can verify that with a system built with cheap devices we obtain clear satellite images.

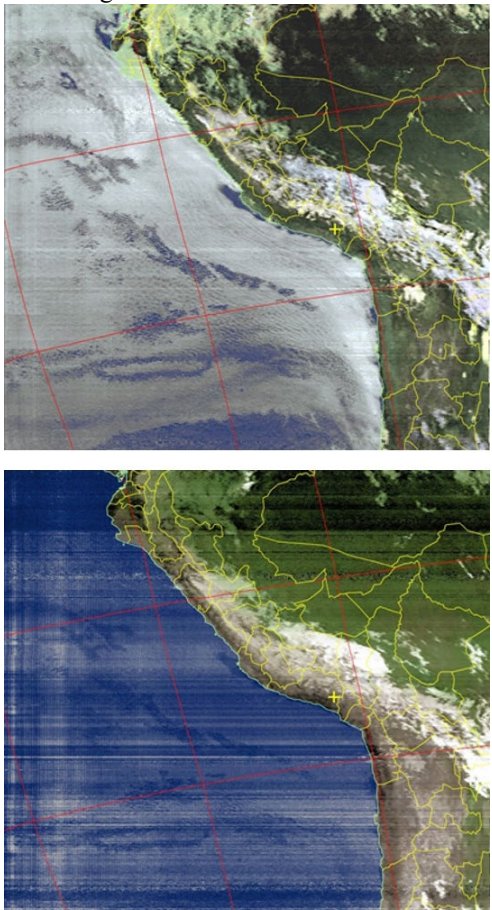

Fig. 8. Image using NOAA 15 signals (upper), the same image but improved (lower).

## IV.

### IV. CONCLUSION

In this work we have developed a system to receive signals from NOAA satellites. For that we have designed and built two types of antennas: a Turnstile crossed dipole antenna with balun and a quadrifilar helicoidal antenna. The RTL SDR RTL2832U receiver was chosen, a device with good qualities and cheap. It is important to choose suitable software for signal processing. Orbitron prediction software is used to detect the passage of the NOAA satellite. The SDRSharp is used to receive and record ljudsignaler. Finally, WxToImg software convert the satellite signals to images and then image process. The results are very promising, the images received are very clear. Now that we have the original images and data, as future work is to section the area to be studied and then use algorithms to study the soil humidity.